\documentclass[twocolumn,showpacs,aps,prl,
amsmath,superscriptaddress,nofootinbib]{revtex4}
\usepackage{amsmath}
\usepackage{epsfig}
\usepackage{epsf}
\usepackage{array}

\newcommand{\beq}{\begin{equation}}
\newcommand{\eeq}{\end{equation}}
\newcommand{\be}{\begin{equation}}
\newcommand{\ee}{\end{equation}}
\newcommand{\beqa}{\begin{eqnarray}}
\newcommand{\eeqa}{\end{eqnarray}}
\newcommand{\bea}{\begin{eqnarray}}
\newcommand{\eea}{\end{eqnarray}}
\newcommand{\eqs}{\end{split}}
\newcommand{\bqs}{\begin{split}}

\newcommand{\nn}{{\bf n}}

\newcommand{\E}{\mbox{\boldmath $E$}}
\renewcommand{\j}{\mbox{\boldmath $j$}}
\newcommand{\nnabla}{\mbox{\boldmath $\nabla$}}

\newcommand{\z}{\mbox{\boldmath $z$}}

\begin{document}

\title{Dynamical symmetry breaking as the origin of the 
zero-$dc$-resistance state in
an $ac$-driven system.}
\author{A.V. Andreev}
\affiliation{Physics Department, University of Colorado, Boulder, CO 80309}
\affiliation{Bell Labs, Lucent Technologies, Room 1D-267, Murray Hill, NJ 07974}
\author{I.L. Aleiner}
\author{A.J. Millis}
\affiliation{Physics Department, Columbia University, New York, NY 10027}
\date{\today}

\begin{abstract}
  Under a strong $ac$ drive the zero-frequency linear response
  dissipative resistivity $\rho _{d}(j=0)$ of a homogeneous state is
  allowed to become negative. We show that such a state is absolutely
  unstable. The only time-independent state of a system with a $\rho
  _{d}(j=0)<0$ is characterized by a current which almost everywhere
  has a magnitude $j_{0}$ fixed by the condition that the nonlinear
  dissipative resistivity $\rho _{d}(j_{0}^{2})=0$.  As a result, the
  dissipative component of the $dc$ electric field vanishes. The total
  current may be varied by rearranging the current pattern
  appropriately with the dissipative component of the $dc$-electric
  field remaining zero.  This result, together with the calculation of
  Durst \emph{et. al.}, indicating the existence of regimes of applied
  $ac$ microwave field and $dc$ magnetic field where $\rho
  _{d}(j=0)<0$, explains the zero-resistance state observed by 
 Mani \emph{et al.}
and Zudov \emph{et al.}.
\end{abstract}

\maketitle

\narrowtext

Recently, two experimental groups~\cite{Mani,Zudov} reported observations of
a novel zero-resistance state in two-dimensional electron systems subjected
to a $dc$ magnetic field and to strong microwave radiation. When no
microwave power is applied, Refs ~\cite{Mani,Zudov} observe a longitudinal
resistivity only weakly dependent on magnetic field, at least for the
relatively  small fields (filling factor $\nu >10$)  applied in the experiment.
However, when a high level of microwave power was applied the resistance
developed a strong oscillatory dependence on the applied magnetic field,  with
oscillation
period controlled only by the ratio of the microwave frequency $\omega $ to
the cyclotron frequency $\omega _{c}$. At low $T$ and in certain field
ranges, the dissipative resistance was observed to \textit{vanish} within
the experimental accuracy.

An important step towards the understanding of these observations was taken
in Ref.~\cite{Yale}, which presented a calculation of the effect of microwave
radiation on the $dc$ linear response conductivity of a two dimensional
electron gas. A crucial result of Ref.~\cite{Yale},
see also Ref.~\cite{Ryzhii} for earlier treatment and
Ref.~\cite{Vavilov}
for detailed analysis,
 is the existence of the
regimes of magnetic field and applied microwave power for which the
longitudinal linear response conductivity is negative, 
\begin{equation}
\sigma _{xx}<0.  \label{eq0}
\end{equation}
However, in the literature so far a precise
connection between a negative linear response conductivity and the
experimental observations has not been presented.

In this Letter we show that Eq.~(\ref{eq0}) by itself suffices to
explain the \emph{zero-$dc$-resistance} state observed in
Refs.~\cite{Mani,Zudov}, independent of the details of the microscopic
mechanism which gives rise to Eq.~(\ref{eq0}). The essence of our
result is that a negative linear response conductance implies that the
zero current state is intrinsically unstable: the system spontaneously
develops a non-vanishing local current density, which almost
everywhere has a specific magnitude $j_0$ determined by the condition
that the component of electric field parallel to the local current
vanishes.  The existence of this instability is shown, under
reasonable assumptions, to imply the observed zero resistance state.

We consider $dc$ transport in a two-dimensional electron gas exposed
to a static magnetic field and to an $ac$ electric field. We assume
that the local $dc$ electric field $\mbox{\boldmath $E$}$ is related
to the local $dc$ -current density $\j $ via
\begin{equation}
\mbox{\boldmath $E$}=\j \rho _{d}\left( \j ^{2}\right) +\left[
\j \times \mbox{\boldmath $z$}\right] \rho _{H},  \label{eq1}
\end{equation}%
where $\mbox{\boldmath $z$}$ is the unit vector normal to the plane of
the system. The crucial quantity in Eq.~(\ref{eq1}) is the
longitudinal (dissipative) resistivity $\rho _{d}\left( \j
  ^{2}\right) $ whose dependence on current determines the physics we
consider. The form of $\rho _{d}(j^{2})$ is determined by parameters
such as the applied magnetic field $B_{app}$ and the frequency
$\omega $ and power $\mathcal{P}_{ac}$ of the $ac$-field, which we do
not explicitly write. Also, to simplify the
discussion we do not consider nonlinear effects in the Hall
resistivity 
$\rho _{H}$. 
This is not crucial for the
zero resistance state; effects of including it in the theory are
discussed briefly at the end of the paper.

\begin{figure}
\epsfxsize=0.4\textwidth
%\vspace*{0.3\textwidth}
\epsfbox{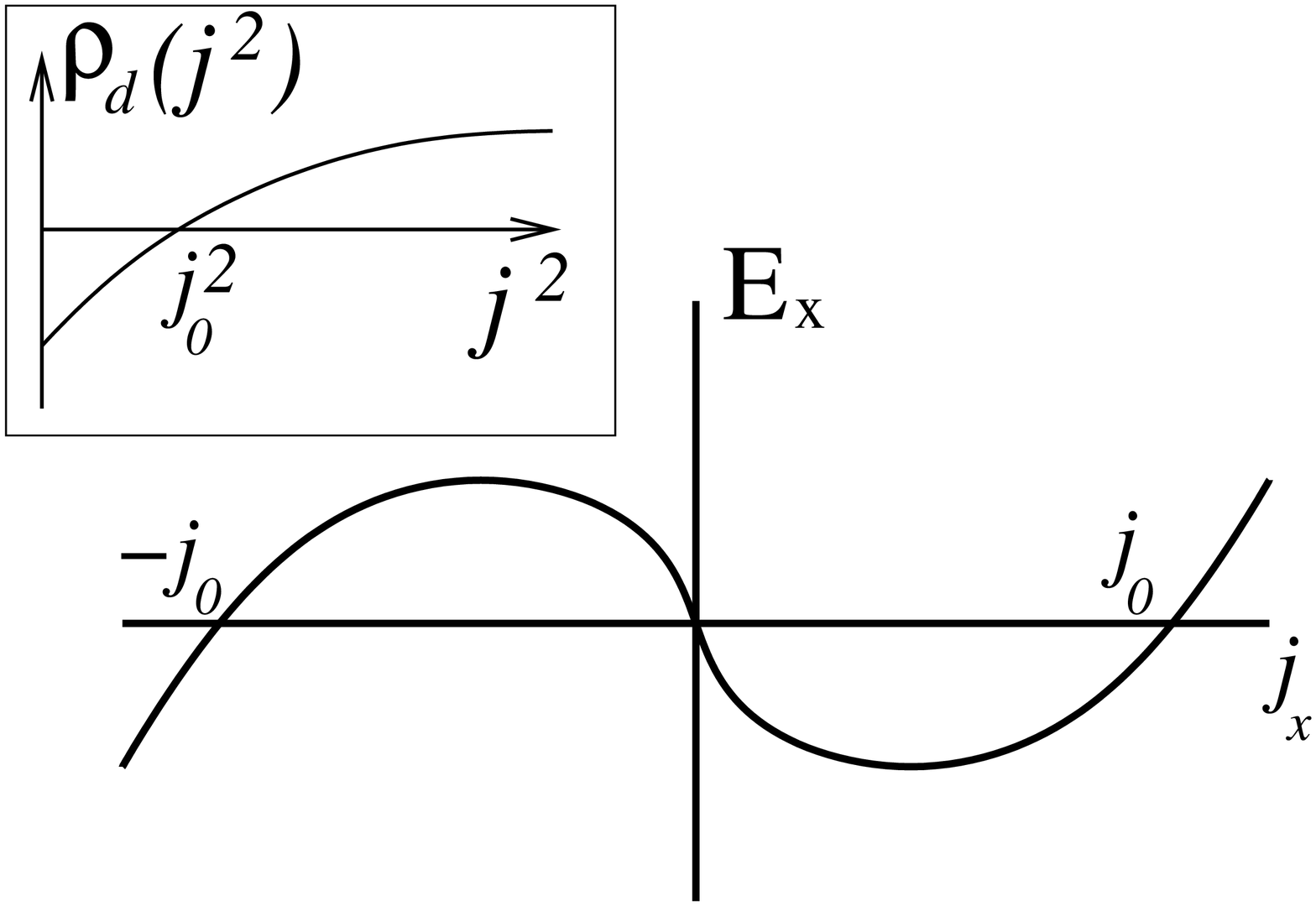}
\caption{Conjectured dependence of the dissipative
component of the local electric field $E_x$ on the current density
$j_x$. The inset shows the current dependence of the dissipative
resistivity.}
\label{fig1}
\end{figure}

We assume that $\rho _{d}\left( \j ^{2}\right) $ is a real,
continuous function of $j^{2}$ and that (as found, e.g. in the
calculations of Ref \cite{Yale}) a range of $B_{app}$, $\omega $ and
$\mathcal{P}_{ac}$ exists for which a spatially homogeneous
zero-current state is characterized by the negative dissipative
resistivity
\begin{equation}
\rho _{d}\left( \j ^{2}=0\right) <0.  \label{eq2}
\end{equation}
However, at sufficiently large values of the $dc$ current $\rho
_{d}(j^{2})$ must revert to its dark ($\mathcal{P}_{ac}=0$) value
because in this limit the microwave radiation will be a small
perturbation on the steady state electron distribution function. 
Continuity  implies that there is a value $j=j_{0}$ at which
\begin{equation}
\rho _{d}(j_{0}^2)=0.
\label{j0def}
\end{equation}
We take $\rho _{d}(j^{2})$ to have the form shown in the inset of
Fig.~\ref{fig1}.
The main panel of Fig.~\ref{fig1} 
 shows the current-voltage characteristic following from the
assumed form of $\rho _{d}(j^{2}).$ Such a dependence was obtained
analytically in Ref.~\cite{Vavilov}.

A negative dissipative resistivity is allowed under non-equilibrium
conditions, if the system is continuously supplied with energy.  In
the situation considered here energy conservation requires only that
$
j^{2}\rho _{d}\left( \j ^{2}\right) +\mathcal{P}_{ac}>0  
$.
However, a negative resistivity may render the system  unstable.
Specifically, we now show that in a system described by
Eq.~(\ref{eq2}) with resistivity curve as shown in the inset in
Fig.~\ref{fig1}.

({\em i}) A homogeneous, time independent state characterized by a current
$j$ of magnitude \textit{ less} than the critical value $j_0$ defined
in Eq.~(\ref{j0def}) is unstable with respect to inhomogeneous current
fluctuations.

({\em ii}) The only possible time independent state is one in which the
current $\j$ has magnitude $j_0$ everywhere except at isolated
singular points (vortex cores) or lines (domain walls), implying vanishing
dissipative electric field, $\j\cdot\E=0$.

An immediate consequence of  ({\em ii}) is that by adjusting the details of
the current pattern, any net $dc$ current less than a threshold value
(which we discuss below) can be sustained at vanishing electric field,
so that \emph{any microscopic mechanism of non-equilibrium drive
  resulting in $\rho _{d}(\j ^{2}=0)<0$ leads to the observed
  ~\cite{Mani,Zudov} zero dissipative differential resistance}:
\begin{equation}
\frac{dV_{x}}{dI_{dc}}=0.  
\label{result}
\end{equation}
(Here $I_{dc}$ is \ a sufficiently weak applied current.) If too large
a current is imposed, the current structure will collapse and a
non-vanishing resistance will be observed. We emphasize, however, that
Eq.~(\ref{result}) is obtained on the assumption that the system is in a
steady state.  Any current pattern consistent with
a boundary condition of small net current implies the existence of
singularities (domain
walls or vortices) in the current distribution; finite density of these
objects may lead to a small dissipative resistivity.

%further at $
%T>0$ such singularities may be thermally created. Motion of these
%objects may lead to a small dissipative resistivity, which we expect
%to vanish rapidly as $T \rightarrow 0$.

Let us pause to discuss the relation of our arguments with
phenomena discussed in the literature.
The instability of systems\cite{Gunn} with absolute
negative conductivity is known since the work of
Zakharov.\cite{Zakharov} The important new feature
of the instability and the domain structure of the present paper is
that it occurs at large Hall angle; as the result
the domains for the current coincide with the domains of the
electric field directed perpendicularly to the current.
We also would like to notice certain similarity with the model
of photoinduced domains proposed by D'yakonov\cite{Dyakonov}
as an explanation of the experiments on Ruby crystals under
the intense laser irradiation\cite{Ruby}.

We now present our specific arguments. We begin by considering the
fluctuations $\delta j$ about a time-independent homogeneous state of
current $j_{i}$. Taking the time derivative of Eq.~(\ref{eq1}) and using the
continuity equation,
\begin{equation}
\frac{\partial n}{\partial t}+\mbox{\boldmath $\nabla$}\cdot \j =0, 
%\tag{\QTO{unrecognized}{\ref{eq1}}$^{\prime }$}  
\label{eq1prime}
\end{equation}
and the Poisson equation,
\begin{equation}
\mbox{\boldmath $E$}=-\mbox{\boldmath $\nabla$}\hat{U}n. 
%\tag{\QTO{unrecognized}{\ref{eq1}}$^{\prime \prime }$}  
\label{eq1primeprime}
\end{equation}
we obtain
\begin{equation}
\mbox{\boldmath $\nabla$}\left( \hat{U}\mbox{\boldmath $\nabla$}\cdot \j
\right) =\frac{\partial }{\partial t}\left[ \j \rho _{d}\left( \j
^{2}\right) +\left[ \j \times \mbox{\boldmath $z$}\right] \rho _{H}
\right] .
\label{eq7}
\end{equation}
Here $n$ is the electron charge density and $\hat{U}$ is a nonlocal
interaction operator which can be expressed in terms of the Green
function of the Laplace equation with appropriate boundary conditions.
The crucial point for us is that $\hat{U}$ has non-negative
eigenvalues.  (We assume that the screening radius is equal to zero,
and neglect the difference between the electric and electrochemical
potentials. This approximation does not alter our conclusions.)

Writing $\j(r,t)=\j_{i}+\delta \j(r,t)$, linearizing in $\delta \j$,
and taking the divergence of both sides of Eq.~(\ref{eq7}), we find
\begin{equation}
\frac{\partial \nnabla \cdot {\delta \j}}{\partial t}=\left( 
{\nnabla}\left(
  \widetilde{\widehat{\rho }}_{d}+\widehat{\mathbf{
\rho }}_{H}\right) ^{-1}\nnabla\hat{U}\right)
\nnabla \cdot \delta \j ,
\label{stability}
\end{equation}
with $\widehat{\mathbf{\rho }}_{H}=\rho _{H}%
\begin{bmatrix}
0 & 1 \\ 
-1 & 0%
\end{bmatrix}
$ the usual Hall resistivity tensor,
\begin{equation}
\widetilde{\widehat{\rho }}_{d}=\rho _{d}\mathbf{1}+
\alpha_j \j_i\otimes \j_i  ,
\label{rhod}
\end{equation}
and 
\begin{equation}
\alpha _{j}=\left. 2\frac{d\rho _{d}(\j ^{2})}{d\j ^{2}}\right\vert
_{\j ^{2}=\j _{i}^{2}}.  \label{alpha}
\end{equation}%
The Coulomb interaction operator $\hat{U}$ is 
positive definite, so the stability is determined
by the sign of the operator  
${\nnabla}\left(
  \widetilde{\widehat{\rho }}_{d}+\widehat{\mathbf{
\rho }}_{H}\right) ^{-1}\nnabla$ in front of it.
Doing Fourier transform of this operator we see that in order
for any solution of Eq.~(\ref{stability}) not to grow with time
the conditions 
\begin{subequations}
\bea
\rho_d(\j^2) \geq 0,
\label{eq12a} \\
\rho_d(\j^2)+\alpha_j\j^2 \geq 0,
\eea
\end{subequations}
must hold.

We therefore conclude that if at least one of $\rho_d$ or
$\rho_d+\alpha_j$, is negative, i.e. if $j_i<j_0$ a homogeneous state
of uniform current is unstable. However, we may also show that any
state with local current density {\em larger} than $j_0$ but  net
current density {\em smaller} than $j_0$ is
necessarily time-dependent.  
In this case the condition $\nnabla \cdot \j =0$
requires the presense of circulating currents.
The integral ${\cal J}=\oint_C d{\bf l}\E$ along the current flow lines
must vanish because $\nnabla \times \E=0$.
On the other hand, from Eq.~(\ref{eq1}) and $\nnabla \cdot \j =0$,
we find ${\cal J}=\oint_C d{\bf l}\j \rho _{d}\left( \j ^{2}\right)$.
By construction of the contour $d{\bf l}\j>0$.  Therefore ${\cal J}=0$ 
can be satisfied together with the stability condition (\ref{eq12a}) only
for  $\rho_d(\j^2)=0$, i.e. for $j=j_0$.

Finally, we examine the stability of general states with $|\vec{j}(r)|=j_0$.
In this case $\rho _{d}=0$ but $\alpha >0$; substitution into Eq,~
(\ref{stability}) and use of Eq.~(\ref{eq1prime}) leads to
\be
\left\{\frac{\partial}{\partial t}+
\left(\nnabla \cdot \left[\j_0\times \z\right]\right)
\left(\left[\z \times\j_0\right]\cdot \nnabla\right)
\frac{\alpha \hat{U}}{\rho_H^2} 
\right\}
 \delta n=0,  
\label{j0stability}
\ee
$\nnabla \cdot \delta\j=-\partial_t\delta n$.
Operator $\hat{U}$ is positive definite, while the operator 
$
\left(\nnabla \cdot \left[\j_0\times \z\right]\right)
\left(\left[\z \times\j_0\right]\cdot \nnabla\right)
$ is Hermitian and  is 
non-negative
because it can be presented as $AA^\dagger$. Therefore,
the state (\ref{j0def}) is not unstable,
except possibly at singular points. The investigation
of the stability of the current pattern in the vicinity of
the singular point requires going beyond the local equations
(\ref{j0stability}) and (\ref{eq1}) and will not be done
in the present paper.

 Moreover, one can see from Eq.~(\ref{j0stability}) 
that all the perturbations
decay in time exponentially with the exception of those for which
$\left[\j_0\times\z\right]\cdot \nnabla \hat{U}\delta n=0$,
i.e. with the electric field directed along $\j_0$.
The physical meaning of these zero-modes is all the perturbations
of the current which keep $\j^2=\j_0^2$ and $\nnabla\cdot \j=0$
(most trivial example of such perturbation is a homogeneous rotation
of vector $\j_0$).
These perturbations have zero eigenvalue and, analogously
to Goldstone modes, are a straightforward consequence of the symmetry
breaking induced by the applied nonlinear drive.

We now consider the physical consequences of our results. We found 
that a non-equilibrium system
which has a negative linear response
resistivity, is unstable to the formation of
a state of non-vanishing local current. Almost everywhere
in the sample  the current
has the magnitude $j_{0}$ at which the dissipative resistivity vanishes, but
the direction must vary so that the net current is consistent
with boundary conditions.  The
current distribution must contain singular regions, of negligible volume
fraction, at which the current takes values different from $j_{0}$. 
The arguments relating to time-independent states given above
may be viewed as showing   that it is impossible to construct
a time-independent singularity structure for distributions involving
currents of magnitude greater than $j_0$, whereas it is possible
if in almost all of the sample the current has magnitude $j_0$.
Just as in the theory of superconductivity the
detailed nature and structure of the singular regions (domain walls, vortex
cores or other structures) presumably depends both on boundary conditions and
on short length scale physics. The question cannot be analyzed within the
quasicontinuum/local response function approach used in this paper, and is
an important topic for future investigation.

\begin{figure}
\epsfxsize=0.45\textwidth
%\vspace*{0.3\textwidth}
\epsfbox{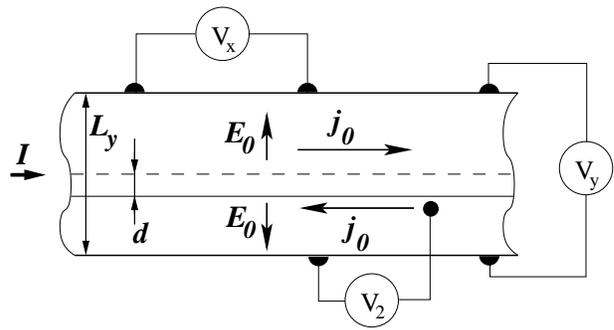}
\caption{The simplest possible pattern of the current distribution
-- domain wall. The net current, $I$, is accommodated by a shift of
the position  of the domain wall by the distance $d$, see text.
Electric field in the domain is $E_0=\rho_{H}j_0$. The current
pattern in the Corbino disc geometry is obtained by
connecting the broken edges into a ring.
}
\label{Fig2}
\end{figure}

For concreteness
of further discussion we will consider the obvious
choice of singularity shown in Fig.~\ref{Fig2}, namely  a linear 
domain wall, separating two
domains in which current flows  parallel and 
antiparallel to the domain wall. We believe
that a structure involving vortices would lead
to essentially identical physics.
In the presence of a magnetic field, consideration of the Hall
component of the current reveals the existence of a discontinuity in the
component of the electric field perpendicular to the boundary. If $\widehat{%
\mathbf{n}}$ is the vector perpendicular to the wall, and $\Delta 
{\j}=2j_{0}$ is the discontinuity in current across the wall
(assumed parallel to the wall direction) then the singularity in the
electric field is 
\begin{equation}
\Delta {\ E}=2\nn\rho _{H}j_{0} . \label{delE}
\end{equation}
This discontinuity requires a charge accumulation which, in a two
dimensional situation, is non-local ($l_{0}$ is a cutoff set by the
microscopic structure of the domain)
\begin{equation}
n(\mathbf{r)}\simeq -\rho _{H}j_{0}\ln 
\left(\frac{\left\vert \mathbf{r\cdot }%
\widehat{\mathbf{n}}\right\vert }{l_{0}}\right).  \label{nofr}
\end{equation}%
This  charge distribution may be detectable by  local probes.
The other possible way to detect dynamical
symmetry breaking is to measure spontaneous voltages
arising inside the domain, voltage $V_2$ in Fig.~\ref{Fig2}.

Figure~\ref{Fig2} presents a very natural 
(albeit probably oversimplified) picture of
the experimental situation studied in Refs \cite{Mani,Zudov}. In these
experiments the current in one direction (say, 'x') was fixed by current
leads to some value $I$, and the current in the transverse direction was
set to zero. The longitudinal (x) and transverse (y) voltages were measured.
Referring to Fig.~\ref{Fig2}, we see that any value of net current $I$
corresponding to a current density much less than $j_{0}$ can be obtained
simply by adjusting the height of the domain wall: if $d$ is the position of
the domain wall relative to the center of the device, then $I=2dj_{0}$
with $V_{x}=0$. Similarly, the total Hall voltage is the sum of a positive
voltage in the upper half of the sample and a negative voltage in the lower
half, leading to $V_{y}=\rho _{H}\left[ j_{0}\left( \frac{L_{y}}{2}-d\right)
-j_{0}\left( \frac{L_{y}}{2}+d\right) \right] =-\rho _{H}I$, which will
equal the dark (no microwave) result if $\rho _{H}$ is not much affected by
the $ac$ field. Notice, that for the Corbino disc geometry the applied
voltage, (corresponding to $V_y$ of Fig.~\ref{Fig2})
 can be also accomodated by the shift of the domain wall
without generation of the dissipative current, resulting in the
{\em zero-conductance-state} \cite{Zudov4}.

The equations analyzed in this paper predict threshold behavior in $I$ at
low temperature $T:$ $V_{x}$ is strictly zero for weak applied currents but
if the applied current is large enough that the current density becomes of
order $j_{0}$ then the domain wall is swept out of the system and a
dissipative state corresponding to current densities greater than $j_{0}$ in
some parts of the sample will result. Similarly, our equations predict a
critical temperature: at very high temperature, the linear response
conductivity will be positive even at non-zero (but fixed) microwave power.
As $T$ is lowered, $\sigma _{xx}$ will decrease and at some temperature pass
through $0$, upon which dissipationless behavior will result. 
The sharp thresholds in $I$ and $T$,
which are in aparent contradition with Refs.~\cite{Mani,Zudov},
may be artifacts of the simple
treatment given here, which assumed a static singularity structure
and zero screening radius.

We did not consider the dependence of the Hall conductivity on
the applied current. It is easy to see that this dependence
does not change the condition (\ref{j0def}) for the circulating
currents in the state because the Hall coefficient does not cause
 dissipation. Singular dependence of $j_0$ on the magnetic field
will cause the singular features in the magnetic field dependence of
the Hall resistivity near the zero-resistance state, The shape
and the value of these singularities, however, have nothing to
do with the quantized plateaux in the Quantum Hall effect.

Finally, we note that we have assumed an isotropic $\rho _{d}$. In fact, the
presence of an $ac$ drive will lead to a quadrupolar 
anisotropy, see Ref.~\cite{Vavilov} for microscopic derivation, 
which for the
sake of notational clarity we did not write but which can easily be included
if desired. This anisotropy will presumably affect the orientations of
domain walls, suggesting that it would be interesting to look for
differences in threshold behavior for $dc$ current parallel or perpendicular
to $ac$ current.

To summarize, we have shown that the remarkable zero resistance state found
by Refs.~\cite{Mani,Zudov} may be understood on very general grounds as a
consequence of a negative linear response conductivity.
Ref.~\cite{Yale} has presented a calculation, based on a
specific microscopic model, showing that this negative linear response
conductivity indeed may occur in the regime of magnetic field and microwave
frequency in which the zero resistance state occurs. Taken together, we
believe the present work and Ref.~\cite{Yale} capture the essence of the
experimental result. 

% Several directions for future research are obvious. On
% the experimental side, we speculate that if instead of imposing a current and
% measuring a voltage, one instead imposes a small voltage and measures a
% current then no steady state can occur. Characterizing the time dependence
% of the resulting situation should lead to interesting insights related to
% domain wall density and motion. On the theoretical side, the crucial
% prediction of our paper is the spontaneous formation of current
% patterns
% characterized by  vanishing dissipation. 
% The precise nature   of the singularities
% implied by the boundary conditions of small net current, 
% have not been determined, nor have the nature of thermally activated
% defects which may lead to the small residual
% resistivity. More generally, a complete set of dynamical equations, going
% beyond the local approximation used here, is needed.

\acknowledgements We are grateful to 
H. Stormer for bringing this problem to our attention and 
and to N. Read, H. Stormer, R. Willett and M. Zudov for valuable
discussions. Two
of us (A.A. and I.A.) were supported by the Packard foundation. A.A. was
supported by the grant NSF-DMR-9984002 and A.J. M acknowledges 
NSF-DMR-00081075.

\end{document}